\documentstyle[aps,psfig,prb,floats]{revtex}
\ifx\undefined\psfig\def\psfig#1{ }\else\fi

\begin{document}
\ifpreprintsty\else
\twocolumn[\hsize\textwidth%
\columnwidth\hsize\csname@twocolumnfalse\endcsname
\fi

%\begin{document}
\draft
\preprint{}

\title{Quantum fluctuations of classical skyrmions in 
Quantum Hall Ferromagnets}
\author{M. Abolfath}
\address{
%Department of Physics, Indiana University,
%Bloomington, IN 47405 \\
Center for Theoretical Physics and Mathematics, 
P.O.Box 11365-8486, Tehran, Iran \\
Institute for Studies in Theoretical Physics and Mathematics,
P.O. Box 19395-5531, Tehran, Iran}

\date{\today}

\maketitle

\begin{abstract}
\leftskip 2cm
\rightskip 2cm

In this article, we discuss the effect of the zero
point quantum fluctuations to improve the results of 
the minimal field theory which has been applied
to study %SMG  the 
skyrmions in the quantum Hall systems.
Our calculation which is based on the semiclassical
treatment of the quantum fluctuations, shows that the 
one-loop quantum correction
provides more accurate results for the minimal field theory.
\end{abstract}


%\pacs{73.40.Hm,73.20.Dx}

\ifpreprintsty\else\vskip1pc]\fi
\narrowtext

The novel development of the semiconductor technology has led to
interesting observations of topological objects, skyrmions,      
in two component quantum Hall systems \cite{Barrett} where
the spin is considered as the additional degree of freedom
in the single layer quantum wells. 
These recent experiments have provided    
a situation in which %SMG that the 
exotic topological field theories could be 
examined.
These models which have been developed
for the elementary particles within the high energy physics, 
have been generalized to make a reliable field theoretic model  
at the limit of low energy physics for condensed   
matter systems, e.g., quantum Hall ferromagnets (QHF). 
Among those theories, the novel idea of the generalized 
Skyrme model \cite{Skyrme} has attracted the attention of
several authors. \cite{Lee,Sondhi,Moon,Fertig,Abolfath}
It has been shown \cite{Sondhi,Abolfath} that 
the accuracy of this model is satisfactory when the Zeeman energy
is small compared to the Coulomb energy between the electrons,
$e^2/\epsilon \ell_0$  where $\ell_0$ is the magnetic length 
and $\epsilon$ is the dielectric constant of the host semiconductor.
Within this approximation, spin is mapped to a
classical unit vector (spin coherent state), as an order parameter.  
This correspondence has been obtained
within the framework of the classical field theory,
where the appropriate mean field theory for the low energy
quantum Hall effect (QHE) has been considered as the 
Chern-Simon-Landau-Ginsburg model (CSLG). \cite{Lee}
Integrating out the charge fluctuations of the composite bosons yields
a minimal non-linear $\sigma$ (NL$\sigma$) model for the spin, 
generalized to include Coulomb
and Zeeman interaction terms. \cite{Sondhi,Moon}
This has led to observing the finite size skyrmion spin texture where
the size of skyrmion is given by the competition between the Zeeman 
energy and the Coulomb energy.
Recently, the range of validity of such minimal field theory 
description has been investigated by 
a comparison with  microscopic Hartree-Fock (HF),
exact diagonalization, and a proposed variational wave function
by Abolfath {\em et al.}. \cite{Abolfath}
It has been shown that the minimal
field theory description is accurate  for
skyrmions with large spin quantum numbers, $K$ ($\gtrsim 10$) 
although, as expected, it fails quantitatively for 
the baby skyrmions ($K < 10$).  
The microscopic HF can be regarded as an effective classical mean field
theory with the appropriate boundary conditions,
which retains the higher order gradient terms
(these terms are absent in the minimal field theory).
On the other hand, the zero point quantum fluctuations (ZPQF)
which are present in the quantum models
(exact diagonalization and the variational wave function),
are absent in these classical mean field theories.
The higher order gradient terms are needed to improve the results
of the minimal field theory to match the microscopic HF level and
the ZPQF's are necessary for the quantum corrections.
Recently Moon and Mullen \cite{Mullen} pointed out for baby skyrmions,
i.e., those typical in GaAs samples \cite{Barrett},
the charge-density interaction within the CSLG theory,
changes from a Coulombic one $1/r$ to $\ln(1/r)$ at short distances.
They have claimed that the minimal field theory with the logarithmic
interaction yields a good quantitative agreement with microscopic study.
In this article we report on a study of the one-loop quantum corrections
of the charged skyrmions in QHF as an alternative approach.
We suggest that the differences between the 
minimal field theory and the microscopic picture 
for the baby skyrmions can be reduced by including the ZPQF.
For instance, the microscopic calculation shows
that the absolute value of the z-component of the spin 
is less than unity at the center of the skyrmion.   
In contrast, within the classical mean field theories 
the spin is constrained to be down at the origin. 
In fact, this is one of the boundary conditions 
used to solve the Euler-Lagrange differential equation.
Such discrepancy can not even be removed by including the
infinite number of higher-order
derivative terms to the minimal field theory and/or replacing the effective
Coulomb interaction between skyrmions
with the logarithmic interaction, the approach which has been 
reported by Moon and Mullen. \cite{Mullen}
The ZPQF modifies the shape of skyrmion,
consistent with the microscopic pictures.
These corrections which are imposed on the classical models, 
are present in any non-colinear spin system, e.g., antiferromagnets.
The negative poles of their Green's function are responsible
for the presence of the ZPQF.
For baby skyrmions, the ZPQF is sufficiently severe %SMG
that the modification of the classical solutions becomes significant.
Conversely, we do not find any correction to the Belavin-Polyakov's solutions
\cite{Polyakov,rajaraman} where the result of the
minimal field theory is identical to the microscopic picture.\cite{Abolfath}
In this case, the ZPQF is negligible for large enough skyrmions
(adult skyrmions).
We finally discuss the importance of the higher order gradient terms.
We start to take the fluctuations into account by making
use of the functional integral approach within the gaussian 
level of approximation.
Note that the skyrmion charge of a class of
configurations can not be changed by local fluctuations.   
The possible way to keep the charge and/or winding number, 
$Q [\equiv \int d {\bf r} \rho ({\bf r})]$, invariant, is:
relaxing the ${\bf m}_z$ due to in-plane fluctuations even at the
boundaries of the system.
Therefore in the space of $Q$'s, ergodicity is broken and 
the path integration must be evaluated for 
a pure state characterized by a given $Q$.
Let us consider the fluctuations around the mean field solution for a 
given $Q$-skyrmion where its functional integral is in the following form:
\begin{eqnarray}
Z_Q = \oint {\cal D} {\bf m} \; \delta ( {\bf m}^2 -1 ) 
e^{ - S[{\bf m}]/ \hbar } \; .
\label{eq7}
\end{eqnarray}
$S$ is the Euclidean action, $S = S_{WZ}[\Gamma] + S_H$.
The first term is the usual Berry's phase of spin history over a closed orbit,
$\Gamma$, on the unit sphere in the presence of 
a unit magnetic monopole at the center,
i.e., the Wess-Zumino term \cite{Fradkin}
and $S_H \equiv \int_0^{\hbar\beta} d \tau E[{\bf m}]$ where
\begin{equation}
E[{\bf m}]=E_0[{\bf m}]+E_z[{\bf m}]+E_{\rm Coul}[{\bf m}] \;.
\label{eq1}
\end{equation}
$E_0[{\bf m}]$ is the leading order term in a gradient expansion
of the energy functional, i.e., the conventional  
Non-linear (NL)$\sigma$-model,
and $E_z[{\bf m}]$ is the Zeeman energy functional.
For QHF where
skyrmions are carrying electric charge, 
next to the leading term
in the gradient expansion is the non-local Coulomb energy, 
$E_{\rm Coul}[{\bf m}]$, 
where the appropriate charge density associated with this term 
is equivalent to the filling factor $\nu$ times the Pontryagian density.
\cite{Sondhi,Moon}
%Within the minimal field theory, 
The static solution of the spin configuration, 
${\bf \tilde{m}}$, satisfy a non-linear differential equation which can be 
obtained by minimizing the energy functional, Eq.(\ref{eq1}), 
with respect to ${\bf m}$.\cite{Abolfath}
%using a Lagrange multiplier to enforce the constraint 
%$\bf m(\bf r)\cdot \bf m(\bf r) = 1$. \cite{Abolfath}
To compute the quadratic fluctuations around this saddle
point solution, we must diagonalize 
the quadratic action. 
%\begin{eqnarray}
%S \; &=& S_c[\tilde{\bf m}] + 
%\; \frac{1}{2} \int d {\bf r} \int d {\bf r}^\prime \;
%\delta {\bf m}_\alpha({\bf r}) \; \nonumber \\&& \times
%\left( \frac{\delta^2 S}{\delta {\bf m}_\alpha({\bf r}) \; 
%\delta {\bf m}_\beta({\bf r}^\prime)} 
%\right)_{\bf m=\tilde{m}} \;
%\delta {\bf m}_\beta({\bf r}^\prime)\; .
%\label{eq9}
%\end{eqnarray}
Following Polyakov, \cite{Polyakov,Polyakovbook}
we exploit the renormalization procedure via 
introducing a complex scalar field, 
$\psi({\bf r}) \; = \; \phi_1({\bf r}) + i  \phi_2({\bf r})$,
where the fluctuation of ${\bf m}$ around its minimal solution is
parametrized by
${\bf m}=\sqrt{1-\overline{\psi}\psi} \; 
{\bf \tilde{m}} \; + \; \sum_{a=1}^{2} \phi_a {\bf e}^a$ and
${\bf \tilde{m}}({\bf r}) \cdot {\bf e}^a({\bf r}) = 0$. 
The ${\bf \tilde{m}}({\bf r})$ defines a map between the compactified 
representation space and the order parameter space. 
It induces a metric and therefore a set 
of connections on the order parameter manifold.
The connections may be interpreted as the geometrical 
Chern-Simon gauge field strength,
$A_\mu \equiv - {\bf e}_1 \cdot \partial_\mu {\bf e}_2 / 2$.
Therefore, the scalar complex field describes a system of  
charged bosonic quasi-particles in the presence of 
%the external magnetic field as well as
the geometrical gauge field.
%field-strength tensor, $F_{\mu\nu} = \partial_\mu A_\nu - \partial_\nu A_\mu$.
\cite{Polyakovbook,Auerbach}
Here we consider a stationary spin-texture hence no time-dependent 
geometrical gauge field strength, 
i.e., a pure geometrical magnetic field.
One may easily show that the geometrical-magnetic field is 
proportional to the spin-texture density (electron density)
\cite{Polyakovbook,Auerbach}, e.g., 
$\nabla \wedge {\bf A}({\bf r}) = \frac{2\pi}{\nu} \rho ({\bf r})$.
This is the zeroth component of the conserved current
which is obtained as 
$J_\mu=(\nu/2\pi)\epsilon_{\mu\nu\lambda}\partial_\nu A_\lambda$
by Lee and Kane in the context of the CSLG theory. \cite{Lee}
The topological charge of skyrmions, $Q$, gives the
vorticity of the geometrical gauge field, $\phi_v = 2\pi Q/\nu$.
Interchanging two of these vortices yields an Aharonov-Bohm phase
factor, $\exp(i \pi Q/\nu)$. This correctly reproduces the original
statistics of the skyrmions.
We interprete the super-current of the
vortices, $J_\mu$, as the topological current of the skyrmions and/or
the quantum Hall current.
Now we may expand the action,
up to the quadratic terms about its classical solution 
($\psi_c=0$) where $\delta {\bf m} = \phi_a {\bf e}^a
- \overline{\psi}\psi \tilde{{\bf m}}/2 + {\cal O}(\phi^4)$ and
%In terms of the scalar complex field, $\psi$, the
%harmonic term is a non-diagonal $2 \times 2$ matrix
%Diagonalizing this matrix, turns out to the following effective action:
%Applying the saddle point approximation,
%the quadratic Wess-Zumino term can be expressed in the
%following manner: 
%\begin{equation}
%S_{WZ} = S_{WZ}[\tilde{\bf m}] +
%\frac{\hbar}{8 \pi \ell_0^2} \int_0^{\hbar\beta} d \tau \int d {\bf r} \;
%\overline{\psi}({\bf r}, \tau)\frac{\partial}{\partial \tau}
%\psi({\bf r}, \tau) \; .
%\label{eq15a}
%\end{equation}
%Up to the gaussian term, the Hamiltonian can be written as 
%$S_H \equiv S_H[\tilde{\bf m}] + \delta^2 S_H$.
%Note that the effective Hamiltonian has two zero eigenvalues corresponding
%to global rotation and translation. \cite{Stone}
%Therefore, one has to treat the zero modes
%properly by making use of the collective coordinates. \cite{rajaraman}
%After a lengthy, but straightforward calculation, one may obtain the
%quadratic term of the effective action
%\begin{eqnarray}
%S_{\rm eff}[\overline{\psi}, \psi] &=& 
%\hbar\beta \sum_n \int d {\bf r} \;
%\overline{\psi}_n({\bf r}) \; \left\{
%\frac{-i\hbar\omega_n}{8 \pi \ell_0^2} \right. \nonumber \\&& \left.
%+ \frac{1}{2}
%\left(\frac{\delta^2 {\cal H}}{\delta \overline{\psi}_n({\bf r}) \; 
%\delta \psi_n({\bf r})} \right)_{\psi=0} \;\right\} 
%\; \psi_n({\bf r}) \; ,
%\label{22}
%\end{eqnarray}
\begin{equation}
S_{\rm eff}[\overline{\psi}, \psi] =  \frac{1}{2}
\int_0^{\hbar\beta} d \tau \int d {\bf r}
\left(\overline{\psi} \; \psi \right) \hat{\cal H}_{\rm eff}
\left( \begin{array}{c} \psi \\ \overline{\psi}
\end{array}\right),
\label{22}
\end{equation}
where
\begin{equation}
\hat{\cal H}_{\rm eff} \equiv\frac{1}{2} \left( \begin{array}{cc}
\frac{\hbar}{8\pi\ell_0^2} \frac{\partial}{\partial \tau}
+\hat{H}({\bf r}) & U({\bf r})  \\
U^\ast({\bf r}) &
\frac{-\hbar}{8\pi\ell_0^2} \frac{\partial}{\partial \tau}
+ \hat{H}^\dagger({\bf r})
\end{array}\right).
\label{22.a}
\end{equation}
In Eq.(\ref{22.a}), the imaginary time derivative term is the
the quadratic Wess-Zumino term and
$\hat{H}({\bf r})=\rho_s (\frac{1}{i} \nabla - 2 {\bf A})^2 + V({\bf r})$
is a second order Hermitian operator and $\rho_s$ is the spin stiffness.
The effective Hamiltonian, $\hat{\cal H}_{\rm eff}$
has two zero eigenvalues corresponding
to global rotation and translation. \cite{Stone}
One has to treat the zero modes
properly by making use of the collective coordinates. \cite{rajaraman}
To evaluate $\hat{H}({\bf r})$, 
we exploit $\partial_\mu A_\mu = 0$, associated with the 
appropriate choice of the Coulomb gauge.
A modified form of $\hat{\cal H}_{\rm eff}$ for the simple NL$\sigma$
model has been presented in Ref. 11. %\cite{Auerbach}.
The off diagonal terms in ${\cal H}_{\rm eff}$ are the remnant of
the Mott-insulating gap for the bosonic fields, $\psi$, where the vortices
superconducting through the system. This can be
interpreted as a dual picture for the quantum Hall state where the
Mott-insulating gap of the bosons and the super-current of the vortices 
are equivalent to the quantum Hall gap and
the quantum Hall current, respectively.
Recently, this duality has been applied to investigate the phase diagram
of the many-skyrmionic quantum Hall systems.\cite{Abolfath-Ejtehadi}
%This is a dual picture for the
%extended skyrmions where $\psi({\bf r})$ are their point-particle partner.
Armed with the second functional derivative form of action,
one may eventually evaluate the contribution of the 
one-loop quantum correction 
to the classical solutions of the quantum 
Hall NL$\sigma$-model.  
The rest of this paper is devoted to the study of 
the local fluctuations associated 
with a single skyrmion ($Q = \pm 1$) where its 
classical solution exhibits the circular symmetry and
$A_\phi = (1 + \tilde{\bf m}_z)/2r$.
%In this case, it is convenient 
%to parametrize the order parameter, ${\bf m}({\bf r})$, by spherical
%field variables. It is obvious that
%and the angular momentum along the $\hat{z}$-axis
%is a good quantum number.
One may expand the complex field, $\psi({\bf r}, \tau)$, 
in terms of the eigenstates of the $\hat{H}$ and
%expanding the temperature-dependent fields in terms of their
its temperature dependent
Fourier transform
\begin{equation}
\psi({\bf r}, \tau) \; = \; \sum_{n, \Omega, m} \; C_{n, \Omega, m}
 \; f_{\Omega, m}(r) \; e^{i m \varphi} e^{-i\omega_n\tau},
\label{23}
\end{equation}
where $\omega_n=2\pi n/\beta$ are the bosonic Matsubara frequencies, 
$f_{\Omega, m}(r)=\langle r | \Omega, m \rangle$
are the orthonormal radial eigenfunctions,  
$\hat{H} |\Omega, m \rangle \; = \; \varepsilon_{\Omega, m}
|\Omega, m \rangle$, $\varepsilon_{\Omega, m}$ is real
and $\langle \Omega', m' |\Omega, m \rangle =\delta_{\Omega',\Omega}
\delta_{m', m}$.
%The global in-plane translation, and the rotation around the 
%$\hat{z}$-axis are denoted by the zero mode energies, 
%$\varepsilon_{\Omega,m}=0$. 
After a little algebraic manipulation,
%and excluding the zero modes from the spectrum of the collective coordinates,  
the effective action can be obtained
\begin{eqnarray}
S_{\rm eff}&&[\overline{C} ,C] = \frac{\hbar \beta}{4}
\sum_n \sum_{\Omega, m} \nonumber \\&& \times
\left(\overline{C}_{n, \Omega, m} \; C_{-n, \Omega, -m}\right)
{\cal S}_{n, \Omega, m}
\left(\begin{array}{c} C_{n, \Omega, m}  \\
\overline{C}_{-n, \Omega, -m}\end{array}\right),
\label{26}
\end{eqnarray}
where
\begin{eqnarray}
{\cal S}_{n, \Omega, m} \equiv \left(\begin{array}{cc}
\frac{-i\hbar\omega_n}{8 \pi} + \varepsilon_{\Omega, m} & U_{\Omega, m} \\
U^\ast_{\Omega, m} &  \frac{i\hbar\omega_n}{8 \pi}
+ \varepsilon_{\Omega, -m}
\end{array}\right).
\label{26.a}
\end{eqnarray}
Here the scale of length is $\ell_0=1$ and $\omega_{-n}=-\omega_n$.
The contribution of the one-loop quantum corrections %fluctuations
to the classical ground state energy,
may be obtained by integrating out the fluctuations $\overline{C}$, 
and $C$. It leads to $E[{\bf m}] = E_c[{\bf \tilde{m}}] + E_f$ 
where $E_c[\tilde{\bf m}]$ is the energy of the classical skyrmion and
\begin{eqnarray}
E_f &=& 
\lim_{\beta \rightarrow \infty} \frac{1}{\beta} \;
\sum_n \sum_{\Omega, m}
%\nonumber\\&& \times
\ln \{ \left(\frac{\beta}{4}\right)^2
[-\left(\frac{i\hbar\omega_n}{8\pi}\right)^2   
+\frac{i\hbar\omega_n}{8\pi}
\nonumber\\&& \times
\left(\varepsilon_{\Omega, m} - \varepsilon_{\Omega, -m} \right)
+ \varepsilon_{\Omega,m} \varepsilon_{\Omega, -m}
 - |U_{\Omega, m}|^2] \}.
\label{27}
\end{eqnarray}
We perform the Matsubara sum in the usual
way to obtain $E_f$.
It turns out an expression for $E_f$, i.e., a sum over the Bose-Einstein
distribution functions, $n_B(\pm \beta\omega^\alpha_{\Omega,m})$. Here
$\omega^\pm_{\Omega,m}=4\pi(\varepsilon_{\Omega, m}-\varepsilon_{\Omega, -m}
\pm \sqrt{(\varepsilon_{\Omega, m}+\varepsilon_{\Omega, -m})^2 -
4 |U_{\Omega, m}|^2})$ are the poles of the action, $\cal S$,
and $\alpha$ stands for either positive or negative sign.
At zero temperature, $n_B(x)$ is
zero or $-1$ for positive or negative $x$'s.
Therefore, the problem of evaluating the fluctuations at zero temperature 
is converted to the problem of finding 
the negative poles of ${\cal S}$. 
For the ferromagnet ground state where all spins are lined
up, $U=0$ and the eigenvalues are continuous and positive
($\omega^\pm \propto k^2$) hence no ZPQF.
However, for the non-colinear spin texture namely
the antiferromagnets, both negative and positive poles
($\omega^\pm \propto \pm k$) are present. At zero temperature,
the negative poles contribute to ZPQF, and
giving the correct shape of the ground state in agreement with the
standard result of the Holstein-Primakov transformation.\cite{Steve}
Note that in any case, ${\cal S}$ is positive definite to guarantee that
the classical solutions are the real minima of the action.
To see this, let us set $\omega_n$ to zero and diagonalize the
matrix in Eq.(\ref{26.a}). We find that eigenvalues are positive. 
%This is also valid for the skyrmions.
In order to estimate the effect of the ZPQF on
the classical skyrmionic solution and then the 
enhancement on its energy, 
we evaluate the spectrum of the ${\cal S}$.
The numerical calculation shows that the $m=0$ is 
the most significant channel which contributes to ZPQF at $T=0$.  
The effective potential of the $\hat{H}({\bf r})$, contains two terms when
the angular momentum, $m$ is non-zero. A linear term, $m A_\phi/r$
and a quadratic term, $m^2/r^2$.
Near the core of skyrmion, 
the linear term is negligible in comparison with the
quadratic term, since $A_\phi \sim r$.
%near the center of the skyrmion, and goes to zero as $1/r$, at infinity.
Approximately, we have $\varepsilon_{\Omega, m}=\varepsilon_{\Omega, -m}$ then
$\omega^\pm_{\Omega,m} = \pm 8\pi
\sqrt{\varepsilon^2_{\Omega, m} - |U_{\Omega, m}|^2}$.
In this case, the $\omega^+_{\Omega,m}$ is positive
and there will be no contribution to ZPQF at $T=0$. We find:
\begin{equation}
E_f = 8\pi \sum_{\Omega, m} 
\sqrt{\varepsilon^2_{\Omega, m} - |U_{\Omega, m}|^2}.
\label{en1.1}
\end{equation}
The effect of the fluctuations is to increase the energy cost of the
skyrmions. $E_f$ (which is proportional to the energy gap)
is positive and decreases as the size of the
skyrmion is increased, i.e., $E_f(K) > E_f(K+1)$.
The energy differences
between a skyrmion with $K$ and $K+1$ spin-flips, $\Delta(K)$, 
estimates the level crossing between two skyrmions, 
as pointed out by Abolfath {\em et al.}. \cite{Abolfath} %SMG
One may note that $\Delta(K)$ can be renormalized by the ZPQF 
\begin{equation}
\Delta(K) = \Delta_c(K) + \Delta_f(K),
\label{en1}
\end{equation}
where $\Delta_f(K) = E_f(K) - E_f(K+1)$,
is the contribution of the ZPQF on $\Delta_c(K)$,
the bare energy differences. \cite{Abolfath}
One may also show that
the value of the Zeeman splitting factor 
($2t$) corresponds to $\Delta(K)$. Then it
is renormalized via ZPQF ($t \rightarrow t^\ast$).
We note that $\Delta_f(K)$ vanishes rapidly as $K$ increases,
when the minimal field theory matches the microscopic results.
We conclude that the use of the leading
gradient terms in the minimal field theory is seriously in error for
small $K$ and the ZPQF can not even give a better prediction for
$\Delta(K)$. This reconfirms the previous results of Ref. 7.
Conversely, including the ZPQF can improve the shape of the skyrmions,
and their $z$-component of the classical solution,
$\tilde{\bf m}_z({\bf r})$.
In the following we study the effect of the ZPQF on ${\bf m}_z(r)$ via
exploiting the above techniques to find out the correct shap of the
skyrmions.
It can be taken into account through 
\begin{equation}
{\bf m}_z ({\bf r}) = {\bf \tilde{m}}_z ({\bf r})
\sqrt{1-{\overline\psi}({\bf r}) \psi({\bf r})}
+ \sum_{a=1}^2 \phi_a({\bf r}) {\hat z} \cdot {\bf e}^a({\bf r}).
\label{28_0}
\end{equation}
We may evaluate the effect of the quantum fluctuations on 
${\bf m}_z$ via the standard technique of gaussian integrals and
power series expansion of $\sqrt{1-{\overline\psi} \psi}$.
Note that the last term in Eq.(\ref{28_0}) vanishes after integration.
Defining the single point correlation function
\begin{equation}
G({\bf r}) = \langle {\overline\psi}({\bf r}) \psi({\bf r}) \rangle
\equiv  \lim_{\tau \rightarrow 0^-}
\langle T_\tau \psi({\bf r}, \tau) {\overline\psi}({\bf r}, 0) \rangle,
\label{29}
\end{equation}
where the ensemble average is denoted by $\langle \dots \rangle$
and $T_\tau$ is the time ordering operator and 
taking advantage of the Wick's theorem turn out the expectation 
value of the spin's z-component in terms of the $G({\bf r})$
\begin{equation}
\langle {\bf m}_z ({\bf r}) \rangle = \Bigl(
1 - \frac{1}{2}G({\bf r})+\frac{1}{4}G^2({\bf r})
\Bigl) \; {\bf \tilde{m}}_z ({\bf r})
+ {\cal O}(G^3).
\label{28_2}
\end{equation}
We can evaluate $G({\bf r})$ after
expanding $\psi$ in terms of the $f_{\Omega, m}(r)$'s. 
Integrating out the fluctuations 
and summing upon the Matsubara frequencies at zero temperature leads to 
\begin{equation}
G({\bf r}) =  -32 \sum_{\Omega, m} f^2_{\Omega, m}(r) \left(
1 - \frac{\varepsilon_{\Omega, m}}{\sqrt{\varepsilon^2_{\Omega, m}
- |U_{\Omega, m}|^2}} \right).
\label{G}
\end{equation}
\begin{figure}
%\centerline{\psfig{figure=m_z.eps,width=1.1\columnwidth}}
\caption{
The radial distribution of spin, $m_z$, 
is plotted. The effect of the 
ZPQF (solid line) on the classical distribution of spin 
(dashed line) is demonstrated for
the $K=1$ skyrmion. 
For comparison, the solution 
of the microscopic HF picture  
(dotted line) is plotted.
The absolute value of $m_z$ at the origion is suppressed by ZPQF.
In the inset, the dashed line and the dotted line show
the effective potential, 
$V(r)$ ($\rho_s$), 
and its bound state (WF), for the $m=0$ channel, respectively. 
%This bound state is responsible for 
%renormalizing the classical skyrmion.
}
\label{Fig2}
\end{figure}
Note that $\hat{H}({\bf r})$ depends on the shape of the
background spin texture, e.g., the size of skyrmion.
For non-zero $m$, the term like $1/r^2$ in $\hat{H}(r)$
smears its dependence on the shape of the skyrmion.
The magnitude of the eigen-energies for the non-zero angular momentums
is large enough that one may neglect their contribution in Eq.(\ref{G}),
i.e., the $m \neq 0$ channels
do not change the magnitude of ${\bf m}_z(r)$ significantly.
%For the sake of simplicity, we may exclude these terms in Eq.(\ref{en1}) to
%capture the physics of the ZPQF on the classical NL$\sigma$-model.
The $m=0$ potential, and its unique bound state ($\Omega = 1$) is shown
in the inset of Fig. \ref{Fig2}. Our choice for the boundary condition
is $df(r)/dr=0$ at the origin.
The curvature of the, 
$V({\bf r})$, is reduced by increasing the size of skyrmion.
For the smaller skyrmions, the depth of the potential
becomes more negative.
The radial distribution of the spin for
the $K=1$ single skyrmion is shown 
in Fig. \ref{Fig2}. 
As one may see, the effect of the ZPQF is severe 
at the center of the skyrmion and
decays at large distances. As we have expected on general 
grounds, the magnitude of the  
spin at the center becomes closer to the predicted in the microscopic 
picture.
%Within this work, we observed that the ZPQF changes
%the shap and the energy of the
%$1/r$-minimal baby skyrmions. Effectively, this is equivalent with a
%picture in which the repulsive range of interaction among the baby
%skyrmions becomes shorter than $1/r$.
%This brings us into qualitative agreement with Ref.{\cite{Mullen}}.

The author thanks S.M. Girvin for the helpful discussion and
inspiring comments on the subject and for providing copies of his
unpublished works and H.A. Fertig, A.H. MacDonald, J.J. Palacios, 
and S. Rouhani for helpful discussions.
The work at Indiana University is supported by NSF DMR-9714055.
I would like to thank Indiana University at Bloomington where most of this
work was carried out.
%%%%%%%%%%%%%%%%%%%%%%%%%%%%%%%%%%%%%%%%%%%%%%%%%%%%%%%%%%%%%%%%%%%%%%%%%%%%%%%


\begin{references}

\bibitem{Barrett}
S.E. Barrett, G. Dabbagh, L.N. Pfeiffer, K.W. West, and R. Tycko,
Phys. Rev. Lett. {\bf 74}, 5112 (1995);
R. Tycko, S.E. Barrett,
G. Dabbagh, L.N. Pfeiffer, and K.W. West, Science {\bf 268}, 1460 (1995);
A. Schmeller, J. P. Eisenstein, L. N. Pfeiffer, and K.
W. West, Phys. Rev. Lett. {\bf 75}, 4290 (1995);
E. H. Aifer, B. B. Goldberg, D. A. Broido,
Phys. Rev. Lett. {\bf 76}, 680 (1996);
V. Bayot, E. Grivei, S. Melinte, M.B. Santos, M. Shayegan,
Phys. Rev. Lett. {\bf 76}, 5484 (1996).


\bibitem{Skyrme}
T.H.R. Skyrme, Proc. R. Soc. {\bf A262}, 233 (1961).

\bibitem{Lee}
D.H. Lee, and C.L. Kane, Phys. Rev. Lett. {\bf 64}, 1313 (1990).

\bibitem{Sondhi} S.L. Sondhi, A. Karlhede, S.A. Kivelson,
and E.H. Rezayi, Phys. Rev. B {\bf 47}, 16419 (1993).

\bibitem{Moon} K. Moon, H. Mori, Kun Yang, S.M. Girvin,
A.H. MacDonald, L. Zheng, D. Yoshioka, and Shou-Cheng
Zhang, Phys. Rev. B {\bf 51}, 5138 (1995). S. M. Girvin and A. H. MacDonald,
in {\em Novel Quantum Liquids in Semiconductor Structures}, edited by
S. DasSarma and A. Pinczuk (wiley, NewYork, 1996).

\bibitem{Fertig}
H.A. Fertig, L. Brey, R. C\^ot\'e, and A.H. MacDonald, Phys. Rev. B {\bf 50},
16419 (1994);
H.A. Fertig, L. Brey, R. C\^ot\'e, A.H. MacDonald, A. Karlhede, and
S.L. Sondhi, Phys. Rev. B {\bf 55}, 10671 (1997).    

\bibitem{Abolfath} M. Abolfath, J.J. Palacious, H.A. Fertig,
S.M. Girvin, and A.H. MacDonald, Phys. Rev. B {\bf 56}, 6795 (1997).

\bibitem{Mullen}
Kyungsun Moon and Kieran Mullen, preprint cond-mat/9707250.

\bibitem{Polyakov} A. M. Polyakov, Phys. Lett. {\bf 59B}, 79 (1975).
J. P. Rodriguez, Phys. Rev. B {\bf 39}, 2906 (1989).
A. Auerbach, B. E. Larson, and G. N. Murthy, Phys. Rev. B {\bf 43},
11515 (1991).

\bibitem{Polyakovbook}
A.M. Polyakov, {\em Gauge Fields and Strings}
(Harwood Academic, New York 1987).

\bibitem{Auerbach}Assa Auerbach, {\em Interacting 
Electrons and Quantum Magnetism} (Springer-Verlag, 1995).

\bibitem{Fradkin}Eduardo Fradkin, {\em Field 
Theories of Condensed Matter Systems} (Addison Wesley, 1991).

\bibitem{rajaraman} R. Rajaraman, {\em Solitons and Instantons}, (North
Holland, Amsterdam, 1982).

\bibitem{Stone}
Michael Stone, preprint cond-mat/9512010.

\bibitem{Steve}
S.M. Girvin, private communication.

\bibitem{Abolfath-Ejtehadi}
R. C\^ot\'e, A.H. MacDonald, Luis Brey, H.A. Fertig, S.M. Girvin,
and H.T.C. Stoof, \prl 78, 4825 (1997);
M. Abolfath, and M.R. Ejtehadi, submitted to \prb.


\end{references}
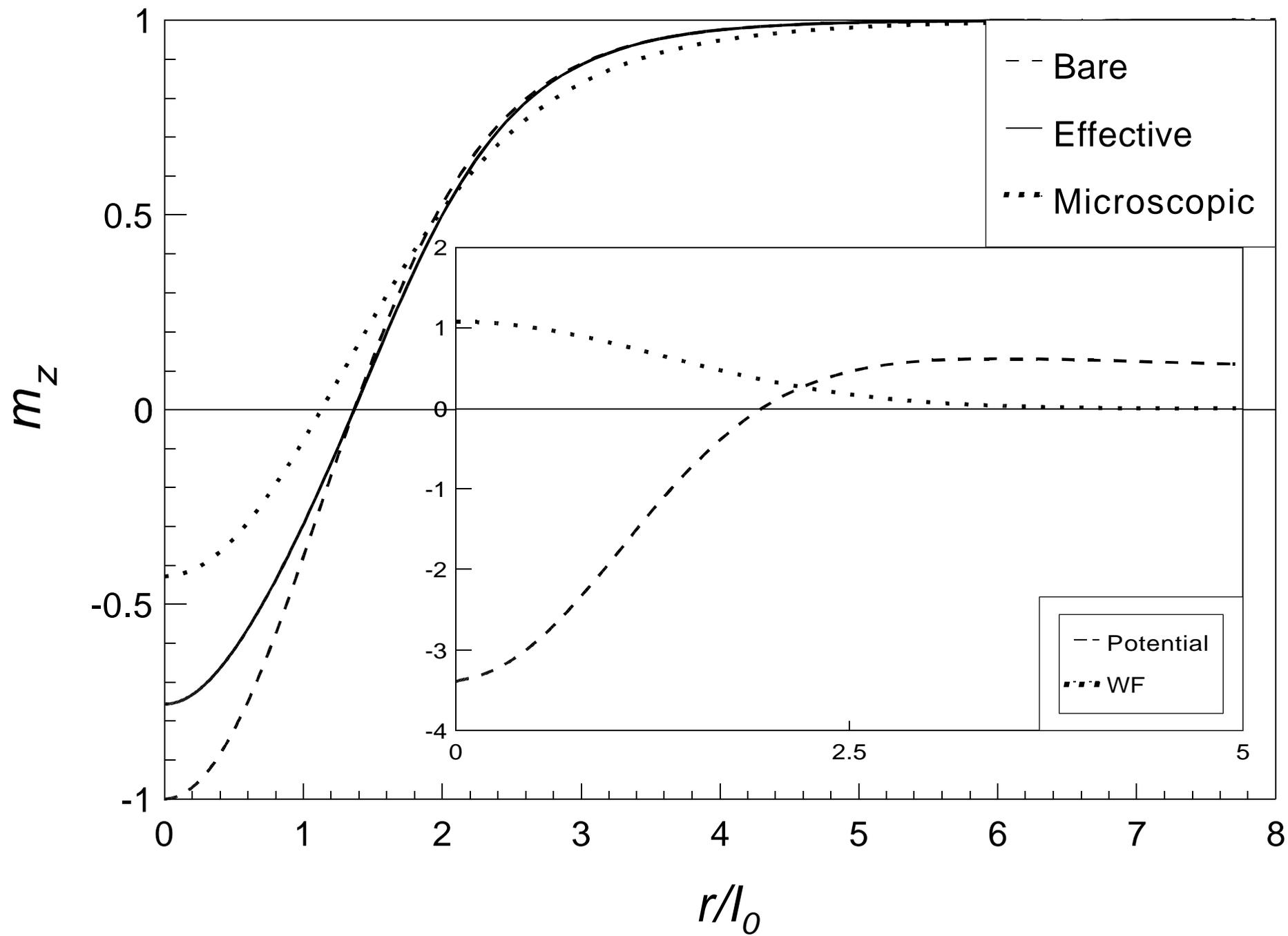
\end{document}